\documentclass[twocolumn]{aastex7}

\renewcommand{\baselinestretch}{1.1}
\usepackage{textcomp}
\usepackage{gensymb}
\usepackage{epstopdf}
\usepackage{amsmath}
\usepackage{graphicx}
\usepackage{hyperref}
\usepackage{multirow}
\usepackage[nameinlink]{cleveref}
\usepackage[multiple]{footmisc}
\crefname{paragraph}{\S}{\S\S} 
\shorttitle{OGLE-2006-BLG-284}
\shortauthors{Terry et al.}

\begin{document}
\title{\textbf{First Direct Identification of a Multi-Star Microlens System Hosting a Planet}}



\author[0000-0002-5029-3257]{Sean K. Terry}
\affiliation{Department of Astronomy, University of Maryland, College Park, MD 20742, USA}
\affiliation{Code 667, NASA Goddard Space Flight Center, Greenbelt, MD 20771, USA}
\email[show]{skterry@umd.edu}

\author[0000-0001-8043-8413]{David P. Bennett}
\affiliation{Department of Astronomy, University of Maryland, College Park, MD 20742, USA}
\affiliation{Code 667, NASA Goddard Space Flight Center, Greenbelt, MD 20771, USA}
\email{bennett.moa@gmail.com}

\author[0000-0003-0014-3354]{Jean-Philippe Beaulieu}
\affiliation{School of Natural Sciences, University of Tasmania, Private Bag 37 Hobart, Tasmania, 70001, Australia}
\email{beaulieu@iap.fr}

\author[0009-0002-6097-9030]{T. Dex Bhadra}
\affiliation{Department of Astronomy, University of Maryland, College Park, MD 20742, USA}
\email{tmbhadra@umd.edu}

\author{Aparna Bhattacharya}
\affiliation{Department of Astronomy, University of Maryland, College Park, MD 20742, USA}
\affiliation{Code 667, NASA Goddard Space Flight Center, Greenbelt, MD 20771, USA}
\email{abhatta5@umd.edu}

\author[0000-0001-5860-1157]{Joshua W. Blackman}
\affiliation{Physikalisches Institut, Universit{\"a}t Bern, Gesellschaftsstrasse 6, CH-3012 Bern, Switzerland}
\email{joshua.blackman@unibe.ch}

\author[0000-0002-8131-8891]{Ian A. Bond}
\affiliation{Institute of Mathematical and Natural Sciences, Massey University, Auckland 0745, New Zealand}
\email{I.A.Bond@massey.ac.nz}

\author[0000-0003-0303-3855]{Andrew A. Cole}
\affiliation{School of Natural Sciences, University of Tasmania, Private Bag 37 Hobart, Tasmania, 70001, Australia}
\email{andrew.cole@utas.edu.au}

\author[0000-0001-7016-1692]{Przemek Mr{\'o}z}
\affiliation{Astronomical Observatory, University of Warsaw, Al.~Ujazdowskie~4,00-478~Warszawa, Poland}
\email{pmroz@astrouw.edu.pl}

\author[0009-0005-3414-455X]{Kansuke Nunota}
\affiliation{Department of Earth and Space Science, Graduate School of Science, Osaka University, Toyonaka, Osaka 560-0043, Japan}
\email{nunota@iral.ess.sci.osaka-u.ac.jp}

\author[0000-0003-2388-4534]{Cl\'ement Ranc}
\affiliation{Sorbonne Universit\'e, CNRS, Institut d’Astrophysique de Paris, IAP, F-75014 Paris, France}
\email{ranc@iap.fr}

\author[0000-0002-1530-4870]{Natalia E. Rektsini}
\affiliation{Sorbonne Universit\'e, CNRS, Institut d’Astrophysique de Paris, IAP, F-75014 Paris, France}
\email{natalia.rec1@gmail.com}

\author[0000-0002-5843-9433]{Daisuke Suzuki}
\affiliation{Department of Earth and Space Science, Graduate School of Science, Osaka University, Toyonaka, Osaka 560-0043, Japan}
\email{dsuzuki@ess.sci.osaka-u.ac.jp}

\author[0000-0002-9881-4760]{Aikaterini Vandorou}
\affiliation{Department of Astronomy, University of Maryland, College Park, MD 20742, USA}
\affiliation{Code 667, NASA Goddard Space Flight Center, Greenbelt, MD 20771, USA}
\email{katievan@umd.edu}

\correspondingauthor{S. K. Terry}

\begin{abstract}

\small \noindent Nearly 25 years have passed since gravitational microlensing delivered its first exoplanet discoveries. The relative proper motion between lens and source systems allows them to be resolved as they separate on the sky over many years. This enables scientists to study light coming directly from the lenses and place tight constraints on the physical properties of these planetary systems. In this paper we present an analysis of high angular resolution images of the microlensing target OGLE-2006-BLG-284 using Keck adaptive optics and HST data taken more than 15 years after the event. We clearly detect the lens system and measure its motion across the Keck and HST epochs. We measure a lens-source relative proper motion of $5.80 \pm 0.15$ mas yr$^{-1}$. A combination of the light curve parameters and the host star(s) brightnesses in $V,\, I$, and $K$ passbands gives primary and secondary host star masses of $M_{\rm 1} = 0.43 \pm 0.05\,M_{\sun}$ and $M_{\rm 2} = 0.12 \pm 0.02\,M_{\sun}$, and a planetary companion mass of $m_p = 165^{+40}_{-33}\,M_{\Earth}$. Although we are unable to determine if the planet orbits one or both stars in the system, our analysis motivates next-generation high-precision RV measurements that can distinguish these two possibilities. This is the first direct detection of flux coming from a two-star microlensing system which hosts a planet.
\\
\\
\textit{Subject headings}: gravitational lensing: micro, planetary systems \\
\end{abstract}


\section{Introduction} \label{sec:intro}
\indent The gravitational microlensing technique \citep{paczynski:1986a, mao:1991a} has a unique detection sensitivity to planets down to 1M$_{\Earth}$ and lower \citep{bennett:1996a}, that orbit their host stars at separations of a few Astronomical Units (AU). The transit and radial velocity methods are insufficient at probing this parameter space. Therefore, microlensing holds an important complementary role in the characterization of exoplanetary systems. It is also important to note that the orbital configuration of the exoplanet does not have to be highly constrained with microlensing as it does with other techniques. Nearly 300 exoplanets have been discovered using the microlensing technique; most of these planets orbit single host stars. This has enabled statistical analyses of this population to be performed \citep{suzuki:2016a, suzuki:2018a, zang:2025a} and tested against current planetary evolution theory \citep{ida:2004a, mordasini:2009a}.\\
\indent Another region of exoplanet parameter space where microlensing has a unique sensitivity is for planets in stellar binary systems with star–star or planet–star separations of a few AU \citep{gould:1992a}. This is the region where microlensing is most sensitive because
it corresponds to the typical Einstein radius ($\theta_E$) for microlensing events toward the Galactic bulge. Kepler \citep{borucki:2011a} has found a number of circumbinary planets in closer orbits around relatively tight binaries \citep{doyle:2011a, welsh:2012a, welsh:2015a}. However, the prevalence of these planets is thought to be a selection effect, as short-period planets are much easier to detect with the transit method. Circumbinary planets in wider orbits, like the first circumbinary planet found by microlensing \citep{bennett:2016a} and the widest orbit circumbinary planet found by Kepler \citep{kostov:2016a}, are thought to form more easily and be more common than circumstellar planets \citep{thebault:2015a}.\\
\indent Of the 19 published triple-lens systems to date, nine of them are reported to have planetary companions in stellar binary systems. With modeling of 1 planet $+$ 2 star systems, it is very difficult to distinguish if the planet orbits both stars in the system (circumbinary) or if the planet orbits a single star in the system (circumstellar). There are several reasons why this interpretation of the orbital configuration is challenging. Firstly, from the light curve modeling alone, only the 2-dimensional instantaneous projected separations (i.e. $a_{\perp}$) between the two stars, the planet and star 1, and the planet and star 2 are estimated. This provides just a lower limit on the instantaneous 3D separation, which itself is only loosely tied to the semi-major axes of the objects in the system. Second, in addition to the classical close/wide degeneracy \citep{griest:1998a, dominik:1999a} and the ecliptic degeneracy \citep{smith:2003a, skowron:2011a}, there are several triple-lens degeneracies that arise from the external-shear approximation, sign reversals on the combined parameters $u_0, \alpha,\phi$ and others \citep{song:2014a, fatheddin:2023a}. Lastly, microlensing parallax and lens orbital motion are more typically unmeasured, particularly for shorter timescale events, but even if orbital motion ($ds/dt, d\alpha/dt$) is measured, these do not give a full Keplerian solution. \\
\indent The one published triple lens system that does have a confirmed circumbinary solution is OGLE-2007-BLG-349 \citep{bennett:2016a}. The authors were able to rule out the circumstellar solution by arguing that the ratio of the separation between the planet$-$center of mass to the star 1$-$star 2 separation is too large to be consistent with a planet orbiting one but not both of the stars. Interestingly, they conclude that this planet does not orbit near the stability limit, where most other circumbinary planets found by \textit{Kepler} reside. We also note the long duration light curve ($t_E\, \sim\, 120$ days) for OGLE-2007-BLG-349 allowed the authors to measure a significant microlensing parallax effect, which resulted in modestly small uncertainties (${\sim}15$\%) on the masses of each lens component. \\
\indent In this paper we re-visit the triple lens microlensing event OGLE-2006-BLG-284. In Section \ref{sec:event} we describe the original ground-based observations. In Section \ref{sec:follow-up} we present the high angular resolution \textit{HST} and Keck adaptive optics (AO) observations and analysis. Section \ref{sec:prop-motion} details our direct measurement of the lens system flux and lens-source separation in the Keck and \textit{HST} data. Sections \ref{sec:light-curve} and \ref{sec:lens-properties} describe our updated light curve modeling which includes constraints from the high-resolution imaging data, as well as our updated interpretation of the lens system properties. Finally, we discuss the overall results and conclude the paper in Section \ref{sec:conclusion}.


\section{Microlensing Event OGLE-2006-BLG-284} \label{sec:event}
OGLE-2006-BLG-284, located at RA $=$ 17:58:38.22, DEC $=$ -29:08:12.0 and Galactic coordinates $(l,b)=(1.277, -2.550)$ was initially alerted by the Optical Gravitational Lensing Experiment (OGLE; \citealt{udalski:1993a, udalski:2015a}). The event was originally published in \cite{jaroszynski:2010a} as a stellar binary event along with other binary microlensing events detected by OGLE between the 2006 and 2008 observing seasons. Although the event was not initially alerted by the Microlensing Observations in Astrophysics (MOA; \citealt{bond01,sumi:2003a}) collaboration, it was recovered as a binary lensing event in the MOA 9-year retrospective project. A joint analysis of the OGLE and MOA photometry was performed by \cite{bennett:2020b} who found a best-fit triple lens model with an improvement of $\Delta \chi^2 \,{\sim}\, 567$ over the previous best-fit binary lens model. Their best-fit solution is a lens system with two stars and a planet. \\
\indent As \cite{bennett:2020b} discuss, the primary difficulty in the interpretation of the planet as circumbinary or circumstellar is the fact that the projected separation between the primary and secondary stars appears to be nearly identical to the projected separation between the planet and the primary star. The authors go on to describe prospects for distinguishing between these two possibilities with future high-resolution imaging of the lens system with Keck AO and next-generation spectrographs on $30\,m-$class telescopes. Successful time-series radial velocity (RV) measurements with ${\sim}2$ km/s precision or better should be able to detect the RV signal of the primary star if it is in a close orbit with the secondary (e.g. $a \lesssim 2$ AU). This would give strong evidence that the two stars are in a close, short-period orbit, and the planet must be in a larger orbit with a large line-of-sight separation (i.e. circumbinary). On the other hand, if there is no significant RV signal measured, this would favor the interpretation that the two stars are in a much wider orbit, and the planet orbits only one star (i.e. circumstellar).


\section{High angular resolution follow-up with Keck \& HST} \label{sec:follow-up}

\subsection{2020$-$2021 $Keck$ Analysis} \label{subsec:keck2020}
\begin{figure*}[!htb]
\includegraphics[width=1.0\linewidth]{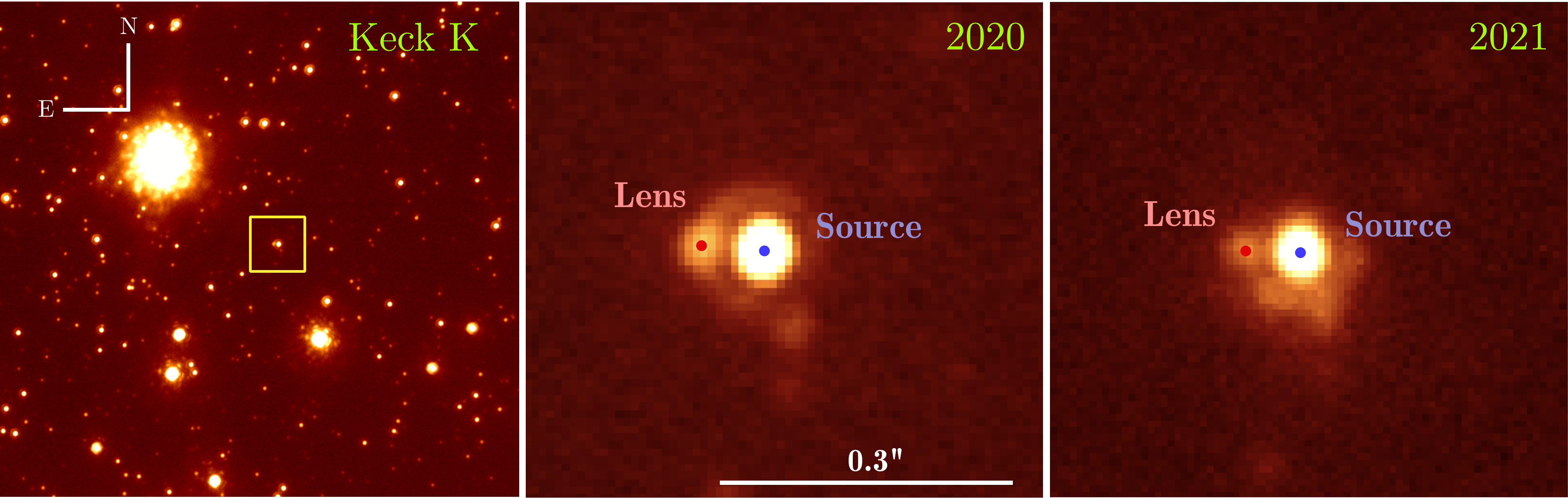}
\centering
\caption{\footnotesize \textit{Left}: Stacked $K$-band image of the OGLE-2006-BLG-284 field from 2020, with the target inside the yellow square. \textit{Middle}: Zoom on the target in the 2020 data. \textit{Right}: Zoom on the target in the 2021 data. The lens and source are clearly resolved, however the 2021 data suffers from noticeable PSF systematics due to degraded AO performance. \label{fig:3panel_keck}}
\end{figure*}

The target OGLE-2006-BLG-284 was observed with the OSIRIS imager on Keck-I in the $K_{prime}$ passband ($\lambda_c = 2.12\mu$m, hereafter $K$) on August 10, 2020 and July 14, 2021. These high-resolution imaging epochs were taken approximately 14.04 and 14.97 years after the peak of the microlensing event in 2006. The 2020 Keck image quality is excellent, with an average point-spread function (PSF) full-width at half-max (FWHM) of 50 mas and Strehl Ratio (SR) of 0.46. The 2021 data is of measurably lower image quality, with an average PSF FWHM ${\sim}$ 60 mas and SR ${\sim}$ 0.26. These two parameters (FWHM, SR) are generally used to understand the performance of the AO system, with smaller FWHM and larger SR values typically indicating better quality. We note there are other, sometimes more subtle, factors that can affect AO image quality that are not necessarily captured by FWHM or SR alone \citep{terry:2023a}. The middle and right panels of Figure \ref{fig:3panel_keck} show a zoom-in on the target(s); the diffraction-limited Airy ring can be seen around the source star in 2020. There appear to be minor systematic effects impacting the AO correction for the 2021 data which lead to lower image quality. The pixel scale for the OSIRIS imager is 9.952 mas/pixel, and all of the images were taken using the Keck-I laser guide star adaptive optics (LGSAO) system. \\
\indent For the 2020 data, a co-add of 25 dithered science images was used for photometric calibration to images from the Vista Variables in the Via Lactea (VVV) survey \citep{minniti:2010a} following the procedure of \cite{beaulieu:2018a}. The images were flat-field and dark current corrected using standard methods \citep{lu:2022a}. We performed astrometry and photometry on the co-added image using SExtractor \citep{bertin:1996a}, and subsequently calibrated the Keck image to VVV by matching 11 bright isolated stars in the frames. This calibration analysis results in uncertainties of 0.05 magnitudes. We select the 2020 dataset to perform the photometric calibration to VVV as it is the higher-quality of the two Keck datasets.

\indent We combined 12 flat-field frames, five dark frames, and 15 sky frames for calibrating our science images. As previously mentioned, a total of 25 science frames with an integration time of 60 s frame$^{-1}$ were reduced using the Keck AO Imaging (KAI) Data Reduction Pipeline \citep{lu:2022a} to correct instrumental aberrations and geometric distortions \citep{lu:2008a, yelda:2010a, freeman:2023a}. Additionally, there were 23 OSIRIS $K-$band images of the target taken on July 14, 2021. Identical reduction and analysis were performed for both Keck epochs. The original motivation to re-visit the target in 2021 was because there were two lens candidates identified in the 2020 epoch; one to the East of the source star and one to the South-Southwest (SSW) of the source star (see middle panel of Figure \ref{fig:3panel_keck}). Both candidates seen in 2020 have separations from the source star that could be consistent with being the lens, given the estimated $\mu_{\textrm{rel}}\, {\sim}\, 6.3$ mas yr$^{-1}$ from \cite{bennett:2020b}. The 2021 data allows us to marginally rule out the SSW object as the true lens, only at the 2$\sigma$-level due to the poorer image quality in 2021. We definitively rule out this SSW object using the 2025 HST $F814W$ data, as we describe in Section \ref{subsec:hst-analysis}.\\
\indent Following the methods of \cite{bhattacharya:2018a} and \cite{terry:2021a}, we use a modified version of the \texttt{DAOPHOT-II} package \citep{stetson:1987a}, that we call \texttt{daophot\_mcmc}\footnote[1]{\href{https://github.com/skterry/DAOPHOT-MCMC}{\scriptsize https://github.com/skterry/DAOPHOT-MCMC}}, to run Markov Chain Monte Carlo (MCMC) sampling on the pixel grid encompassing the targets. Further details of the MCMC routine are given in \cite{terry:2021a, terry:2022a}. We simultaneously fit a multiple-star PSF model to the source and lens (and SSW star) in both 2020 and 2021 datasets, which allows us to extract precise photometry and astrometry while taking into account any effect from their overlapping PSF profiles. Table \ref{table:dual-phot} shows the calibrated magnitudes for the source and lens systems of $K_{\textrm{Source}}=17.42 \pm 0.05$ and $K_{\textrm{Lens}}=19.47 \pm 0.08$. The uncertainties are derived from the ``MCMC+Jackknife method" described in \cite{terry:2021a}. Querying the \cite{Surot2020} extinction map at this location, we find a $K$ band extinction of $A_K = 0.13 \pm 0.05$. From our re-analysis of the light curve modeling (Section \ref{sec:light-curve}), we find a source color of $V_S - I_S = 1.84$, which leads to an extinction-corrected color of $V_{S0} - I_{S0} = 0.71$. We use the color-color relations of \cite{kenyon:1995a} and the source I-band magnitude from Table \ref{tab:lcpar}, $I_S = 19.99$ to predict a source $K$ band magnitude of $K_S = 17.55$. The PSF fit source brightness in Keck is $0.13$ mag brighter than the predicted $K-$band source brightness. Considering the intrinsic uncertainty of ${\sim}5$\% in our adopted color-color relation, combined with the 0.05 mag uncertainty from our Keck measurement, we conclude that nearly all the flux we measure at the position of the source star is due to the source itself. This small excess may suggest a very faint companion to the source star or flux from an unresolved background star.

\begin{deluxetable*}{llll}[!htb]
\deluxetablecaption{HST and Keck PSF photometry\label{table:dual-phot}}
\tablecolumns{4}
\setlength{\tabcolsep}{26.0pt}
\tablewidth{\linewidth}
\tablehead{
\colhead{\hspace{-1.7cm}Star} &
\colhead{$V$ Mag} & \colhead{$I$ Mag} & \colhead{$K$ Mag}
}
\startdata
Lens & $25.51 \pm 0.29$ & $22.91 \pm 0.18$ & $19.47 \pm 0.08$\\
Source & $21.61 \pm 0.19$ & $19.88 \pm 0.10$ & $17.42 \pm 0.05$\\
Lens $+$ Source & $21.58 \pm 0.16$ & $19.82 \pm 0.08$ & $17.27 \pm 0.05$\\
\enddata
\tablenotetext{}{\footnotesize{\textbf{Note}. $V$ and $I$ magnitudes are calibrated to the OGLE-III system and $K$ magnitudes are calibrated to the VVV system as described in section \ref{sec:follow-up}.}}
\end{deluxetable*}

\begin{figure*}[!htb]
\includegraphics[width=0.9\linewidth]{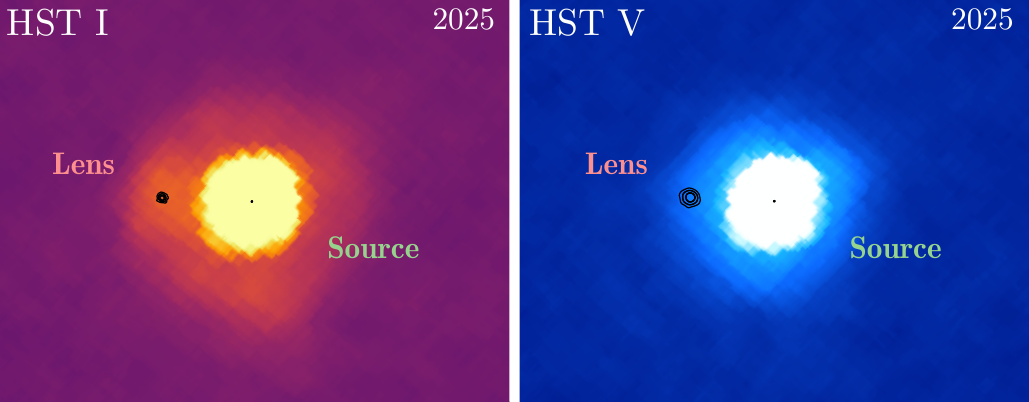}
\centering
\caption{\footnotesize \textit{Left}: The HST $I-$band observation of the target taken in 2025. \textit{Right}: HST $V-$band observation taken during the same 2025 epoch. Contours give the 68\%, 95\%, and 99\% confidence intervals for the lens and source positions. The images show 100 $\times$ 100 super-sampled pixels where the observed dither offsets are accurate to 0.01 pixel. \label{fig:2panel_hst}}
\end{figure*}

\subsection{2025 $HST$ Analysis}\label{subsec:hst-analysis}
OGLE-2006-BLG-284 was observed with HST on 22 April, 2025 as part of program GO-17834 \citep{terry:2024a}. We obtained two orbits of observations using the WFC3/UVIS camera, with 16 $\times$ 69 second exposures taken in the $F814W$ (wide $I-$band) filter and 16 $\times$ 70 second exposures taken in the $F606W$ (wide $V-$band) filter. We reduced the data and performed the PSF fitting analysis using \texttt{MORIA} \citep{bhadra:2026a}, which is a modified version of the original fitting codes presented in \cite{bennett:2015a} and \cite{bhattacharya:2018a}. This procedure includes pixel-level analysis of each individual exposure without any resampling. This avoids any potential loss of resolution that drizzled or stacked, under-sampled images would introduce. \\
\indent Figure \ref{fig:2panel_hst} shows a zoom-in of the target(s) from the HST $I-$band filter (left panel) and $V-$band filter (right panel). The MCMC distributions of source and lens positions are shown as the black contours, and the image orientation is identical to Keck (North-up, East-left). The HST data were obtained approximately 18.74 years after the peak microlensing magnification in 2006. Perhaps more importantly, the HST epoch occurs approximately 4.70 years after the best-quality Keck epoch (2020). This enables us to compare the lens-source separations directly measured in the high-resolution datasets, confirm the lens-source relative proper motion, and definitively rule out the previously mentioned SSW nearby object as the lens. We perform simultaneous three-star PSF fitting of the source$+$lens$+$SSW object in the HST data. Tables \ref{table:dual-phot} and \ref{table:epoch_seps} give the calibrated HST magnitudes, lens-source separations, and lens-source relative proper motion ($\mu_{\textrm{rel,H}}$). Similarly to the calibrated $K-$band source magnitude from Keck, the calibrated $V$ and $I-$band source magnitudes from HST are $0.1-0.2$ mag brighter than the light curve source magnitudes reported in \cite{bennett:2020b}. We again attribute this to a potential faint companion in the source system or unresolved background star light.\\
\indent As previously mentioned, we are able to definitively rule out the SSW star as the lens by measuring the Keck$-$to$-$HST motion. During this 2020$-$2025 time baseline the separation between the source star and the SSW star changes by $\Delta r = -16.4\,\pm\,3.9$ mas, which means the source and SSW star have gotten closer to each other on the sky over time. Therefore we can strongly rule out (at the $11\sigma$ level) the SSW star as the lens.

\subsubsection{Lens System Color-dependent Centroid Shift}\label{CDCS_lens}
\begin{figure*}
\centering
\includegraphics[width=0.75\linewidth]{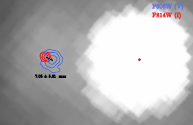}
\centering
\caption{The fitted source and lens system positions as measured in the HST F606W ($V-$band; blue) and F814W ($I-$band; red) data. The contours show the 68\%, 95\%, and 99\% confidence intervals for the $V$ and $I$ data, respectively. The measured $V-$band and $I-$band position of the (brighter) source star is nearly identical. We detect a weak (${\sim}1.3\sigma$) color-dependent centroid shift at the location of the lens system.
\label{fig:cdcs}}
\end{figure*}

\indent \cite{bennett:2006a} and \cite{bennett:2007a} first conceptualized the color-dependence of the flux-weighted centroid in highly blended source and lens star pairs. Given measurably different colors, there should be an apparent difference in the location of a fitted PSF centroid as measured in two different passbands. \cite{bhattacharya:2023a} confirmed the first color-dependent centroid shift measurement made by \cite{bennett:2006a} by independently resolving the OGLE-2003-BLG-235 lens and source stars in Keck AO followup imaging 15 years after the microlensing magnification.\\
\indent We have clearly resolved and individually measured the source star and lens system in both Keck and HST datasets. However, the two stars that reside in the lens system are not individually resolved. Therefore we investigate if there is an appreciable difference in the PSF fit centroid between the HST $I-$band (F814W) and HST $V-$band (F606W). Since the color-dependent centroid shift measurement must be made using PSF fit centroids obtained at approximately the same time, we do not include any Keck $K-$band astrometric measurements in this test as the Keck data were obtained several years before the HST data.\\
\indent We perform this analysis using the results of the two-star (source$+$lens) simultaneous fitting that was described in Section \ref{subsec:hst-analysis}. As Figure \ref{fig:cdcs} shows, there is evidence of a shifted position in the fitted lens system centroids from HST $V$ and $I$ bands. We measure a $\Delta r_{\textrm{V-I}} = 7.05 \pm 5.61$ mas. The difference in each of the North and East components can also be verified by inspection of Table \ref{table:epoch_seps}. The contours shown in Figure \ref{fig:cdcs} are the 68.3\%, 95.5\%, and 99.7\% confidence intervals for the lens system position in $F606W$ (blue) and $F814W$ (red). Clearly the centroid position is less precisely measured in the $F606W$ passband, which primarily contributes to the large error on this measurement. We consider this a marginal detection of a color-dependent centroid shift, at the 1.3$\sigma$ level. As we just mentioned, simultaneously fitting for the position of the source star must be included, otherwise we risk contamination of our astrometric measurement of the lens system flux which could bias the filter-dependent measurements. The fitted source system centroids are shown on the same Figure \ref{fig:cdcs} panel, where the blue$-F606W$ and red$-F814W$ contours nearly fully overlap.\\
\indent Given our results from image-constrained light curve modeling, described in Sections \ref{sec:light-curve} and \ref{sec:lens-properties}, we investigate the implications of this marginal color-dependent centroid shift measurement made in the 2025 HST data. With masses of $M_{\rm 1} = 0.43 \pm 0.05\,M_{\sun}$ and $M_{\rm 2} = 0.12 \pm 0.02\,M_{\sun}$, we use mass-luminosity relations of \cite{delfosse:2000a} and \cite{benedict:2016a} to estimate a flux ratio of 0.01 in $V-$band and 0.04 in $I-$band between the two stars in the lens system. This gives $\Delta f_1 = 0.028$, which is the difference in the primary star's flux fraction between $V$ and $I$ bands (i.e. $\Delta f_1 = \frac{f_{1,V}}{f_{1,V}+f_{2,V}} - \frac{f_{1,I}}{f_{1,I}+f_{2,I}}$). Using the relation $\frac{\Delta r_{\textrm{V-I}}}{\Delta f_1}$ along with the estimated lens system distance of 4.5 kpc, and \cite{bennett:2007a}, we estimate a 2D semi-major axis $a_{\perp,\textrm{2025}} = 231 - 2035$ AU. This large range corresponds to a weakly bound pair of stars in a very-wide orbit. Since the centroid shift measurement itself is only weakly constraining, we consider this weak evidence of the widely-separated binary scenario. We would require nearly a factor of ${\sim}10$ better precision on the HST astrometry in order to place reasonably confident bounds on the projected semi-major axis of the lens system stars.

\section{Lens-Source Relative Proper Motion} \label{sec:prop-motion}
The 2020 and 2021 Keck-I follow up observations were taken 14.04 and 14.97 years after peak magnification in 2006, and the 2025 HST follow up observations were taken 18.74 years after peak magnification. We analyze all three high-resolution epochs independently, derive the per-epoch relative proper motion between the lens and source, and then take the weighted mean of these measurements; $\sigma_{\bar{x}} = \left(\sum_i \frac{1}{\sigma_i^2}\right)^{-1/2}$. We measure the lens-source relative proper motion to be $\mu_{\textrm{rel},H} = (\mu_{\textrm{rel,H,E}},\mu_{\textrm{rel,H,N}}) = (5.77 \pm 0.14, 0.49 \pm 0.03)$ mas yr$^{-1}$, where `H' indicates that these measurements were made in the Heliocentric reference frame, and the `E' and `N' subscripts represent the East and North directions respectively.\\
\indent Since light curve modeling is typically performed in the Geocentric reference frame that moves with the Earth at the time of the event peak, we must convert between the Geocentric and Heliocentric frames. We use the relation derived in \cite{dong:2009b}, $\mu_{\textrm{rel,H}} = \mu_{\textrm{rel,G}} + \frac{{\nu_{\Earth}}{\pi_{\textrm{rel}}}}{AU}$, where $\nu_{\Earth}$ is Earth's projected velocity relative to the Sun at the time of peak magnification. For OGLE-2006-BLG-284 this value is $\nu_{\Earth \textrm{E,N}} = (29.007, 0.351)$ km/sec = $(6.119, 0.074)$ AU yr$^{-1}$ at HJD$' = 3899.47$. Using the expression above and our directly measured $\mu_{\textrm{rel,H}}$ from Keck and \textit{HST}, we determine a relative proper motion in the geocentric frame of $\mu_{\textrm{rel,G}} = 5.26 \pm 0.24\,$mas/yr. There is slight tension between our measured value and the proper motion estimate reported by \cite{bennett:2020b}, however the two measurements agree within $2\sigma$. We discuss possible reasons for this slight tension in Section \ref{subsec:source_companion}.

\begin{deluxetable*}{rccc}[!htb]
\deluxetablecaption{Measured Lens-Source Separations from Keck and HST\label{table:epoch_seps}}
\tablecolumns{3}
\setlength{\tabcolsep}{24.0pt}
\tablewidth{\linewidth}
\tablehead{
\colhead{}&
  \multicolumn{3}{c}{Separation (mas)} \\
\cline{2-4}
\colhead{Year} &
\colhead{East} & \colhead{North} & \colhead{Total}
}
\startdata
$2020\, (Keck\,\, K)$ & $83.04 \pm 4.98$ & $6.89 \pm 1.26$ & $83.33 \pm 5.14$\\
$2021\, (Keck\,\, K)$ & $89.41 \pm 6.14$ & $6.54 \pm 1.76$ & $89.65 \pm 6.39$\\
$2025\, (HST\,\, I)$ & $109.88 \pm 2.52$ & $9.66 \pm 0.59$ & $110.30 \pm 2.59$\\
$(HST\,\, V)$ & $102.98 \pm 4.78$ & $7.43 \pm 1.42$ & $103.25 \pm 4.98$\\
\hline
\hline
 & $\mu_{\textrm{rel,H,E}}$(mas/yr) & $\mu_{\textrm{rel,H,N}}$(mas/yr) & $\mu_{\textrm{rel,H}}$(mas/yr)\\
 \hline
weighted mean & $5.77 \pm 0.14$ & $0.49 \pm 0.03$ & $5.80 \pm 0.15$\\
\enddata
\end{deluxetable*}

\begin{figure*}
\includegraphics[width=0.8\linewidth]{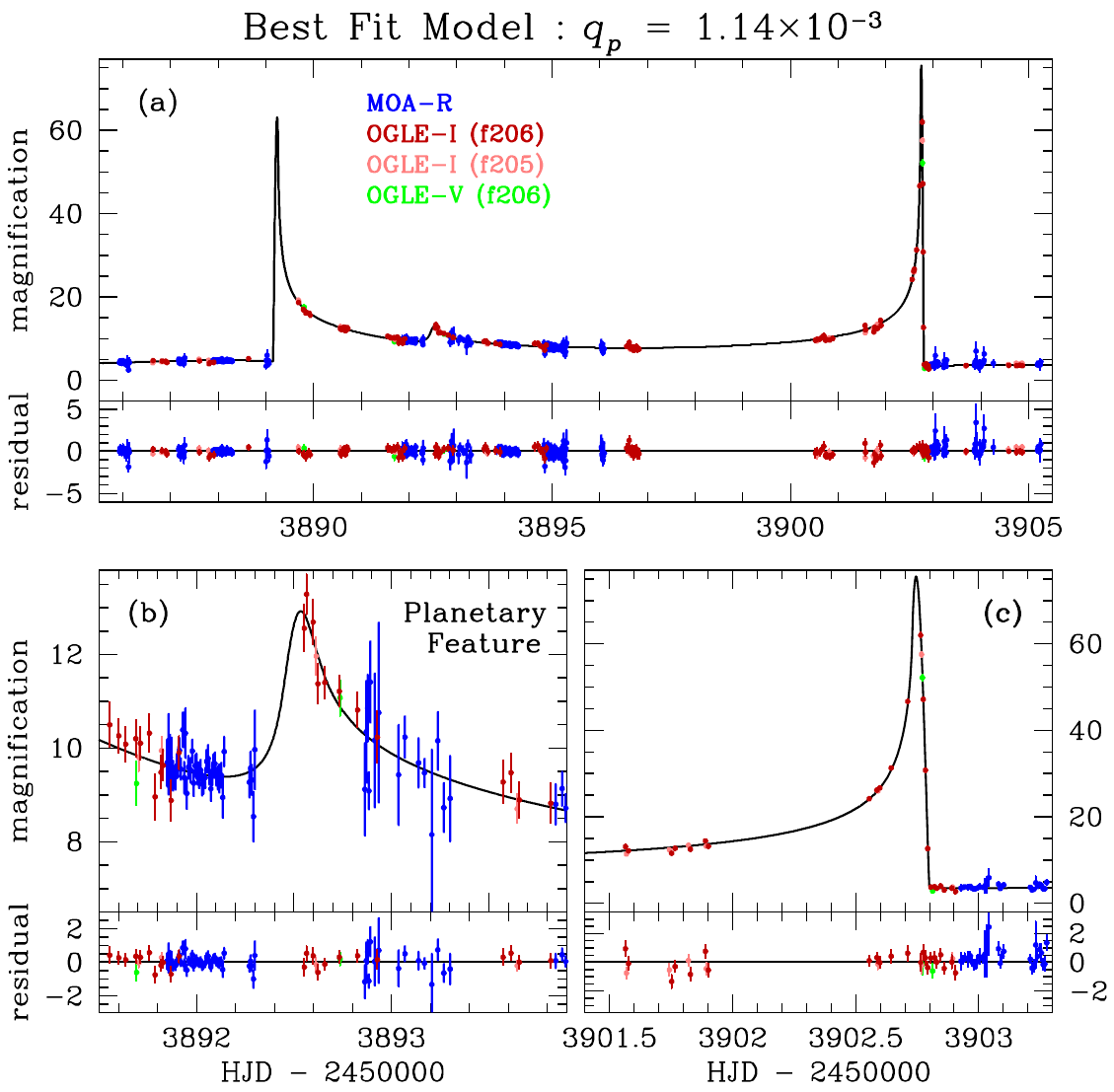}
\centering
\caption{\footnotesize Best fit planetary light curve model for OGLE-2006-BLG-284 with the data used for the analysis in this paper. The upper panels are presented in units of magnification, $A(u) = \frac{u^2 + 2}{u\sqrt{u^2 + 4}}$, and the bottom panels show the residuals to the best-fit triple-lens model. The data behind the figure is available in machine readable format at the journal. \label{fig:lc}}
\end{figure*}

\begin{figure}
\centering
\includegraphics[width=\linewidth]{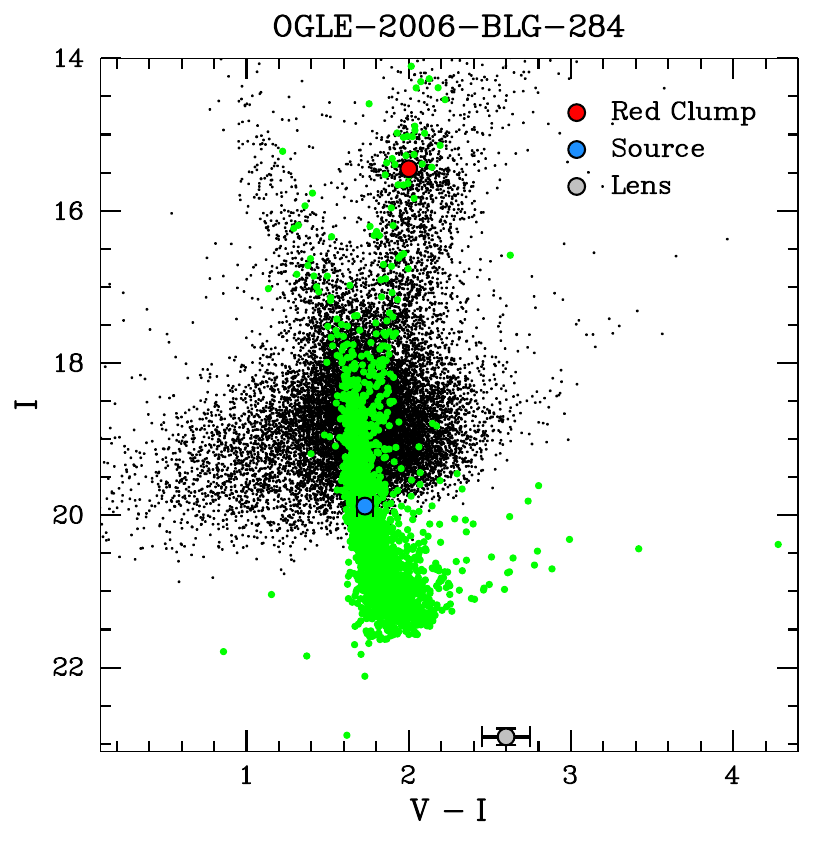}
\centering
\caption{The observed color-magnitude diagram (CMD) for the OGLE-2006-BLG-284 field. The OGLE-III stars within $150^{\prime\prime}$ of the target are shown in black, with the \textit{HST} CMD of all detected sources shown in yellow. The red point indicates the location of the red clump centroid, and the blue and grey points show the source and lens colors and magnitudes from the \textit{HST} observations.
\label{fig:CMD}}
\end{figure}

 
\section{Image-Constrained Light Curve Model} \label{sec:light-curve}
For our light curve modeling, we use the package \texttt{eesunhong} \citep{bennett:1996a}, modified to include constraints on the brightness and separation of the lens and source stars from the high resolution imaging \citep{bennett:2024a}. This procedure helps prevent the light curve modeling from exploring areas in parameter space that are excluded by the high resolution follow-up observations. The search for triple-lens models follows the approach of \cite{bennett:2020b}; first the best stellar binary lens models are determined using an initial condition grid search that replaces the Einstein radius crossing time, $t_E$, and time of closest approach between the lens center-of-mass and the source star, $t_0$, with the times of the primary caustic entry and exit. Then we fix these binary lens parameters and search for triple lens models that can account for the additional photometric feature, as displayed in the lower-left panel of Figure \ref{fig:lc}. Table \ref{tab:lcpar} gives the best fit and MCMC averages for our image-constrained light curve modeling, we find a planetary mass ratio (planet-to-primary star) of $q_2 = 1.14 \times 10^{-3}$ and a stellar binary mass ratio of $q_1 = 0.28$. These are largely consistent with the values reported in the \cite{bennett:2020b} discovery paper.

\begin{deluxetable*}{lcccr}[t]
\deluxetablecaption{Best Fit OGLE-2006-BLG-284 Model Parameters\label{tab:lcpar}}
\tablecolumns{5}
\setlength{\tabcolsep}{18pt}
\tablewidth{\columnwidth}
\tablehead{
\colhead{Parameter} &
\colhead{Units} & \colhead{Best Fit} & \colhead{MCMC Averages} & \colhead{\hspace{6mm}\cite{bennett:2020b}}
}
\startdata
$t_E$ & days & $40.257$ & $40.225 \pm 1.141$ & $39.38 \pm 0.97$\\
$t_{0}$ & HJD$'$ & $3899.469$ & $3899.465 \pm 0.073$ & $3899.425 \pm 0.066$\\
$u_0$ & {} & $0.076$ & $0.076 \pm 0.002$ & $0.078 \pm 0.002$\\
$s_{1\, \textrm{cm}}$ & {} & $0.793$ & $0.795 \pm 0.010$ & $0.801 \pm 0.009$\\
$s_{23}$ & {} & $0.766$ & $0.765 \pm 0.007$ & $0.762 \pm 0.006$\\
$\alpha_{1\, \textrm{cm}}$ & {rad} & $0.322$ & $0.322 \pm 0.004$ & $0.324 \pm 0.004$\\
$\phi_{23}$ & {rad} & $0.057$ & $0.058 \pm 0.011$ & $0.061 \pm 0.009$\\
$q_1$ & {} & $0.282$ & $0.282 \pm 0.013$ & $0.289 \pm 0.011$\\
$q_2$ & {10$^{-4}$} & $11.4$ & $11.39 \pm 1.63$ & $12.6 \pm 1.9$\\
$t_*{}$ & {days} & $0.034$ & $0.034 \pm 0.001$ & $0.034 \pm 0.001$\\
$I_s{}$ & {} & $19.992$ & $19.994 \pm 0.037$ & $19.995 \pm 0.036$\\
$V_s{}$ & {} & $21.834$ & $21.838 \pm 0.038$ & $21.803 \pm 0.036$\\
$\chi^2 / \textrm{dof}$ & {} & $12424.29/12418$ & {} & {}
\enddata
\tablenotetext{}{\footnotesize{\textbf{Note.} The triple lens parameter definitions are as follows: $s_{1\, \textrm{cm}}$ is the separation between mass 1 and the center of mass of masses 2 and 3, $s_{23}$ is the separation between masses 2 and 3, $\alpha_{1\, \textrm{cm}}$ is the angle between the source trajectory and the axis connecting mass 1 with the center of
mass of masses 2 and 3, and $\phi_{23}$ is the angle between this axis (mass 1 and cm of 2 and 3) and the line connecting masses 2 and 3.}}
\end{deluxetable*}

\indent To measure our new lens system parameters, we sum over the MCMC results using a Galactic model \citep{bennett:2014a} with weights for the microlensing rate and our weighted $\mu_{\textrm{rel,H}}$ value from Keck$+$HST (described in Section \ref{sec:prop-motion}). Additionally, we include the source distance as a fitting parameter in the re-modeling of the light curve with imaging constraints. We include a weighting from the \cite{koshimoto:2021a} Galactic model as a prior for $D_S$, and we also use the same Galactic model to obtain a prior on the lens distance for a given value of $D_S$. This prior is not used directly in the light curve modeling, but instead is used to weight the entries in a sum of Markov chain values.
\subsection{Investigating a Source Companion} \label{subsec:source_companion}
During our analysis efforts we noticed that the light curve modeling prefers a slightly fainter source star. As mentioned earlier, the calibrated HST $V$, $I$, and Keck $K$ band source star magnitudes are marginally brighter ($0.1-0.2$ mag) than the source magnitudes given by non-image-constrained light curve modeling. We attributed this to the possibility that there may be a faint companion to the source star that is not magnified significantly. Further, a source companion may also explain the modest discrepancy between our inferred $\mu_\textrm{rel,G}$ from the high-resolution imaging and the $\mu_\textrm{rel,G}$ reported by \cite{bennett:2020b}. If the source has a binary companion that is too widely separated to have a measurable effect on the light curve, it could have a measurable effect on the high-precision astrometry (given a flux ratio ${\sim}\,1$). If the projected separation of the companion is 80 AU this would give an offset of 10 mas between the source and its companion, which is then multiplied by the flux ratio. However, as described in Section \ref{sec:follow-up} and shown in Figure \ref{fig:cdcs}, we do not measure any color-dependent centroid shift at the position of the source system in HST. These two scenarios effectively work to counteract each other.\\ 
\indent Nevertheless, we perform additional light curve modeling that includes a companion to the source star. One version of our analysis assumes blending in the HST images, so that the HST limits on the source are only upper limits on the $V$ and $I$ band source brightness (e.g. lower limits on the source $V$ and $I$ magnitudes). This improves $\chi^2$ by 26.52. An additional version of our modeling includes a source companion (and thus a lower limit on the $V$ and $I$ magnitudes from HST), but it also considers the best caustic crossing solution, which is disfavored by $\Delta\chi^2$ = 22.51. We weight this with a Gaussian probability for this $\chi^2$ difference, which means that it effectively contributes very little. Ultimately, the light curve parameters for the best fitting two-source models are nearly indistinguishable from the best fitting single-source models. For clarity we report only the global best fit model (3L1S) in Table \ref{tab:lcpar}.

\subsection{Source Color \& CMD} \label{source_properties}
Determining the angular size of the source star ($\theta_*$) allows one to estimate the source radius crossing time ($t_*$) through the following relation:
\begin{equation} \label{eq:tstar_equation}
    t_* = \frac{\theta_*}{\mu_{\textrm{rel}}}.
\end{equation}
\noindent But first we need to determine the extinction corrected source magnitude and color. The extinction to the source is found by determining the centroid of the red clump using stars within $150^{\prime\prime}$ of the OGLE-2006-BLG-284 target coordinates. Figure \ref{fig:CMD} shows the observed color-magnitude diagram (CMD) for the OGLE-2006-BLG-284 field. Yellow data points correspond to stars measured in the HST data, and black data points are OGLE-III stars within $150^{\prime\prime}$ of the target. The locations of the red clump, source star, and lens star(s) are shown as red, blue, and grey data points respectively.\\
\indent Following \cite{nataf:2013a}, we determine $I$ and $V$ band extinctions of $A_I = 1.10$ and $A_V = 2.03$. Using the source magnitudes from Table~\ref{table:dual-phot}, we find extinction corrected magnitudes of $I_{S0} = 18.78$  and $V_{S0} = 19.58$. This allows us to use the surface brightness relation from the analysis of \citet{boyajian:2014a}, but we use the following custom formula \citep{aparna16} using stars spanning the range in colors that are relevant for microlensing events:

\begin{equation}
    \textrm{log}({2\theta_{*}}) = 0.5014 + 0.4197(V_{S0} - I_{S0}) - 0.2 I_{S0}
\end{equation}

\noindent This yields $\theta_{*} = 0.603 \pm 0.021\, \mu$as, which is slightly larger than the \cite{bennett:2020b} value of $\theta_{*} = 0.585 \pm 0.029\, \mu$as, however they agree within the error bars.

\begin{figure}
\centering
\includegraphics[width=0.98\linewidth]{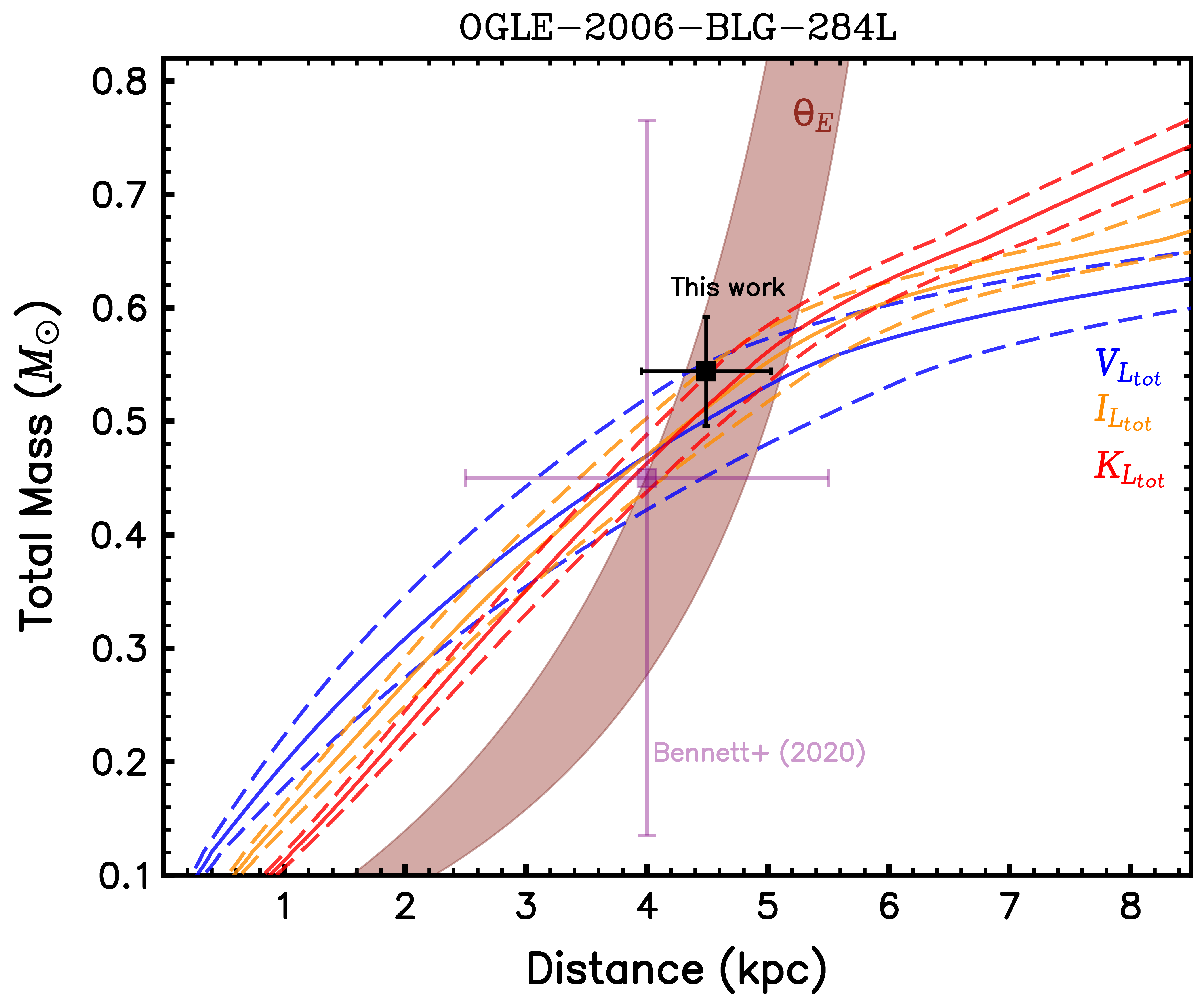}
\centering
\caption{The mass–distance relation for OGLE-2006-BLG-284L with constraints from the total lens flux measurement in HST V (blue), HST I (orange), and Keck K (red). Dashed lines show the 1$\sigma$ error bars for each passband. The solid grey curve shows the mass–distance relation calculated using the angular Einstein radius measurement ($\theta_E$).
\label{fig:md_relation}}
\end{figure}

\section{Lens System Properties} \label{sec:lens-properties}
The angular Einstein radius, $\theta_E$, gives a relation that connects the lens system mass to the source and lens distances, $D_S$ and $D_L$ \citep{bennett:2008b, gaudi:2012a}. The relation is given by:

\begin{equation} \label{eq:md_thetaE}
    M_{L} = \frac{c^2}{4G}\theta_{E}^{2}\frac{D_{S}D_{L}}{D_{S}-D_{L}},
\end{equation}

\begin{figure*}
\centering
\includegraphics[width=0.9\linewidth]{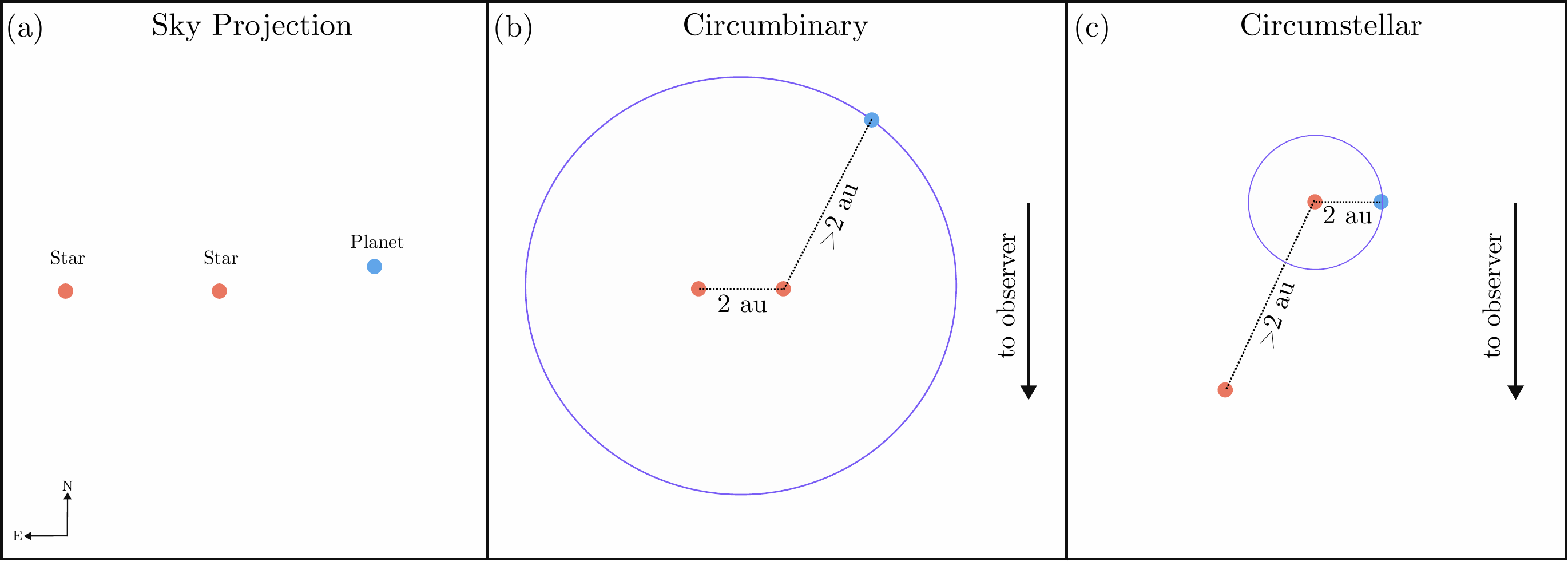}
\centering
\caption{Illustration of two possible orbital configurations for the OGLE-2006-BLG-284L planetary system. Object sizes are not drawn to scale. Panel (a) shows the on-sky projection of the three lens masses (as in Figure 3 of \cite{bennett:2020b}). Panel (b) shows the circumbinary configuration, where the planet is in a wide orbit around both stars. Panel (c) shows the circumstellar configuration, where the planet orbits only one star in the system. Future RV measurements by next-generation ELTs may be able to distinguish between the two scenarios.
\label{fig:orbits}}
\end{figure*}

\noindent where $M_{L}$ is the lens mass, $G$ and $c$ are the gravitational constant and speed of light. As mentioned previously, the measurement of $\mu_{\textrm{rel,H}}$ from the high resolution imaging allows us to measure $\mu_{\textrm{rel,G}}$ to high precision, which ultimately lets us determine $\theta_E \sim \mu_{rel,G} \times t_E$. Figure \ref{fig:md_relation} shows the total mass and distance for the lens system from our analysis in this work (black data point). The solid red, orange, and blue curves show the constraint from the mass-luminosity relation derived from the lens flux measurement in the Keck $K$ band, HST $I$ band, and HST $V$ band imaging, respectively. The dashed lines show the measurement error from the lens flux detections combined with the intrinsic uncertainty in the adopted mass-luminosity relations \citep{delfosse:2000a}. Additionally, the mass-distance relation obtained from the measurement of $\theta_{E}$ (i.e. Equation \ref{eq:md_thetaE}) is shown as a solid dark grey region. For reference we plot the inferred lens system mass and distance reported by \cite{bennett:2020b} as the light purple data point in Figure \ref{fig:md_relation}. Note we plot the total lens system mass in Figure \ref{fig:md_relation}. While the two stars in the lens system do have their own individual mass-luminosity relations, what we directly measure in the Keck and HST data is the combined flux of the system.\\
\indent Table \ref{tab:planet-params} gives the derived lens system physical parameters with RMS errors and 2$\sigma$ ranges. We find that the lens system primary star has a mass of $M_{\textrm{L1}} = 0.426 \pm 0.046M_{\Sun}$, the secondary star has a mass of $M_{\textrm{L2}} = 0.120 \pm 0.015M_{\Sun}$, and the planetary companion has a mass of $M_{\textrm{p}} = 165^{+40}_{-33}\, M_{\oplus}$. We can calculate the semi-major axes between the two stars, as well as the planet and both stars separately. Using the separation values reported in Table \ref{tab:lcpar} and the expression $a_{\perp} = sD_{L}\theta_{E}$, we find a 2D separation of $a_{\perp,s1s2} = 2.03 \pm 0.25$ AU for the stellar separation, $a_{\perp,s1p} = 1.96 \pm 0.24$ AU for the planet$-$Host 1 separation, and $a_{\perp,s2p} = 3.99 \pm 0.51$ AU for the planet$-$Host 2 separation. \\

\begin{figure*}
\centering
\includegraphics[width=0.9\linewidth]{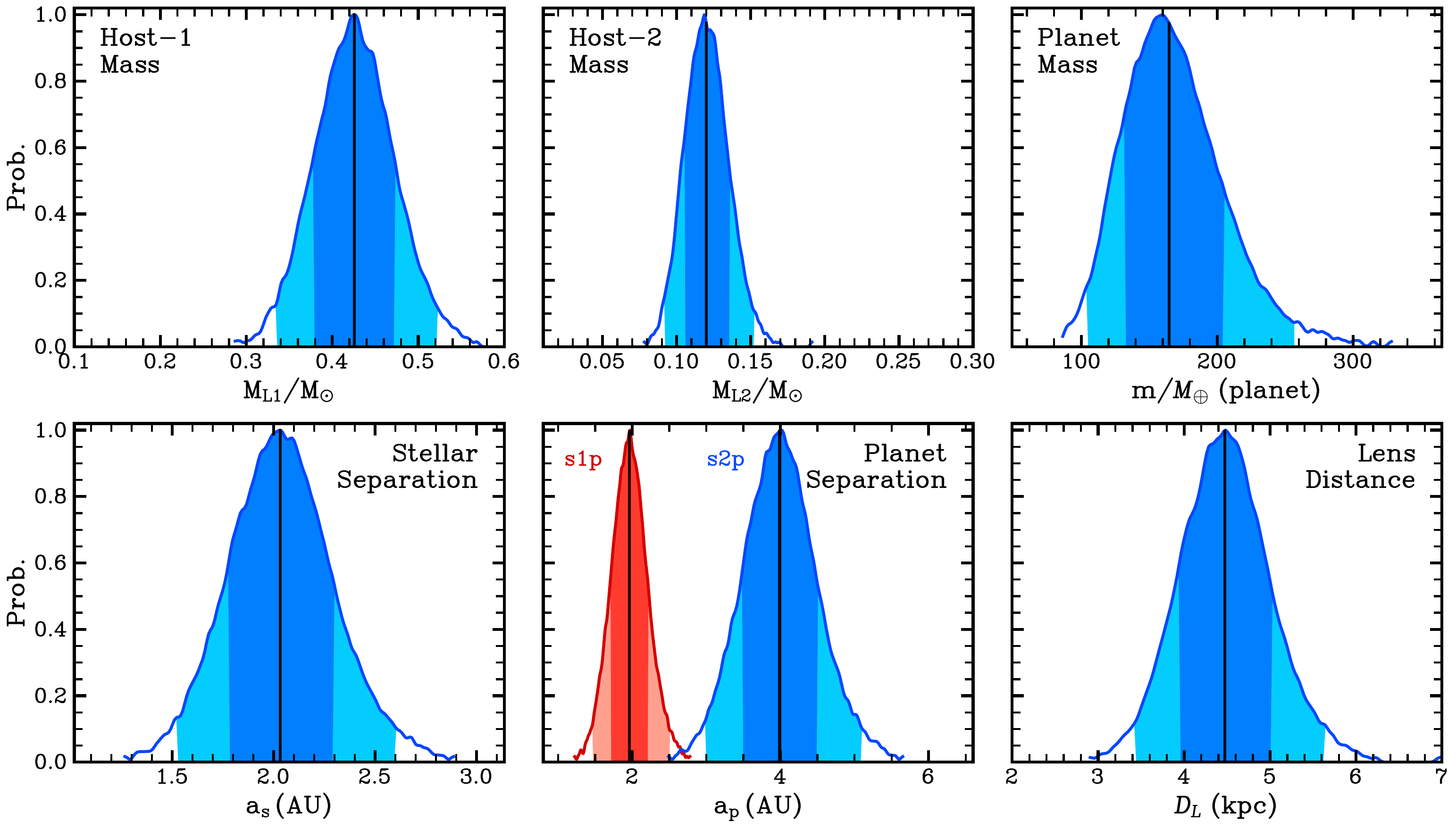}
\centering
\caption{The posterior probability distributions for the lens system physical parameters: primary host mass (upper left), secondary host mass (upper center), planet mass (upper right), stellar separation (lower left), planet-primary host separation (lower center), and lens system distance (lower right). The central 68\% of each distribution is shown in dark blue, with the remaining central 95\% shown in light blue. The vertical black line shows the median of the probability distribution for each parameter.
\label{fig:posteriors}}
\end{figure*}

\subsection{Circumbinary or Circumstellar Planet}\label{subsec:planet-orbit}
About two dozen confirmed circumbinary planets have been discovered to date, mostly from the Kepler \citep{borucki:2011a} and TESS \citep{ricker:2015a} missions. Conversely, several hundred circumstellar planets orbiting one star of multiple-star systems have been discovered\footnote{\url{https://exoplanetarchive.ipac.caltech.edu/}}. The reason for this disparity is both a selection effect and an astrophysical effect. Short period exoplanets are easier to detect with the transit method, and the presence of a binary companion likely interferes with circumbinary planet formation. The inner disk, which contains material that would otherwise be used to form planetesimals, is truncated for circumbinary planets \citep{thebault:2015a}.\\
\indent Although we have made a direct identification of flux coming from the OGLE-2006-BLG-284 triple lens system, we are unable to determine if the planet orbits both stars or a single star in the system. The marginal detection of a color-dependent centroid shift between HST $I$ band and $V$ band data may imply a large separation for the two stars in the system, thus supporting the circumstellar interpretation for the planet. However this $1.3\sigma$ detection is not significant, and at best only places a limit on the projected separation of the two lens stars in 2025. Figure \ref{fig:orbits} panel (a) shows the on-sky projection of the lens system (as in \cite{bennett:2020b}). Panels (b) and (c) of Figure \ref{fig:orbits} show the circumbinary and circumstellar interpretations, respectively. One component of the triple lens system must have a true 3D separation larger than 2 AU from the primary host star. This must be either the lower-mass secondary star or a widely orbiting planetary companion. \\
\indent A time-series radial velocity (RV) measurement of the lens system would provide the strongest evidence yet in support of either remaining scenario. However this measurement is effectively impossible with current-generation spectrographs on large-aperture telescopes. The $K\sim 19.5$ lens system, in very close proximity to the $K \sim 17.4$ source star, is too challenging for current-generation spectrographs on $8-10m$ class telescopes to achieve the necessary RV precision. We expect a next-generation class of extremely large telescope (ELT) spectrographs should have the necessary combination of angular resolution, collecting area, and spectral resolution to attempt direct RV measurements of the lens system and distinguish between the circumstellar and circumbinary scenario. Precisions of $\sim1$ km/s on the faint target should be able to detect the RV signal of the $0.426\, M_{\odot}$ primary star if it has a small semi-major axis with the companion star (circumbinary panel of Figure \ref{fig:orbits}). This would imply that the two lens stars are in a close orbit and the planet must be the third body at a much wider separation. On the other hand, if there is no RV signal detected, this would favor the circumstellar configuration where the primary and secondary stars have a wide orbit and the planetary companion orbits one star only. Finally, the eccentricity and inclination of the lens system is currently not known, which means the expected RV signal could vary from as little as 0.89 km/s at $i = 10\degree$ (nearly face-on) to a maximum of 3.44 km/s at $i = 90\degree$ (edge-on). Since we already have well-measured masses for both stars in the system, the classical `sin i degeneracy' \citep{hilditch:2001a, ho:2011a} is of little concern for our particular purpose.

\begin{deluxetable*}{lccc}[!htp]
\deluxetablecaption{Planetary System Properties from Lens Flux Constraints\label{tab:planet-params}}
\tablecolumns{4}
\setlength{\tabcolsep}{17.5pt}
\tablewidth{\columnwidth}
\tablehead{
\colhead{\hspace{-6cm}Parameter} & \colhead{Units} &
\colhead{Values \& RMS} & \colhead{2-$\sigma$ range}
}
\startdata
Angular Einstein Radius ($\theta_E$) & mas & $0.57 \pm 0.02$ & $0.54-0.60$\\
Geocentric lens-source relative proper motion ($\mu_{\textrm{rel,G}}$) & mas/yr & $5.26 \pm 0.24$ & $4.75-5.74$\\
Host 1 mass ($M_{L1}$) & $M_{\Sun}$ & $0.426 \pm 0.046$ & $0.334-0.518$\\
Host 2 mass ($M_{L2}$) & $M_{\Sun}$ & $0.120 \pm 0.015$ & $0.090-0.150$\\
Planet mass ($M_{\rm p}$) & $M_{\Earth}$ & $165^{+40}_{-33}$ & $99-245$\\
Stellar Separation ($a_{\perp,s1s2}$) & au & $2.03 \pm 0.25$ & $1.53-2.53$\\
Planet$-$Host 1 Separation ($a_{\perp,s1p}$) & au & $1.96 \pm 0.24$ & $1.47-2.46$\\
Planet$-$Host 2 Separation ($a_{\perp,s2p}$) & au & $3.99 \pm 0.51$ & $2.97-5.01$\\
Lens Distance (D$_{L}$) & kpc & $4.47 \pm 0.53$ & $3.41-5.53$\\
Source Distance (D$_{S}$) & kpc & $7.40 \pm 0.98$ & $5.44-9.36$\\
\enddata
\end{deluxetable*}


\section{Discussion and Conclusion} \label{sec:conclusion}

\indent Our high resolution follow up observations of the microlensing target OGLE-2006-BLG-284 have allowed us to make the first ever measurement of flux coming directly from a microlens system comprising two stars and a planetary companion. We measure the total flux of the system in multiple passbands ($V$, $I$, $K$), and make a precise determination of the amplitude and direction of the lens-source relative proper motion $\mu_{\textrm{rel}}$. The lens system flux measurements enable us to use mass-luminosity relations and new constraints on the light curve modeling to measure a precise mass and distance for the lens system. The lens system consists of two main sequence stars with a planetary companion in the disk at a distance from Earth of $D_L = 4.47 \pm 0.53$ kpc. The primary star has a mass of $0.426 \pm 0.046\, M_{\odot}$, the secondary star has a mass of $0.120 \pm 0.015\, M_{\odot}$, and the planet has a `sub-Jupiter' mass of $165^{+40}_{-33}\, M_{\oplus}$ ($\sim0.52M_{\textrm{Jup}}$). Figure \ref{fig:posteriors} shows the posterior distributions for each physical parameter of the lens system with constraints from the high resolution imaging. \\
\indent Using the HST $V$ and $I$ band data taken in 2025, we attempt the first color-dependent centroid shift measurement on a microlens system directly. This has been successfully demonstrated before on a blended source-lens pair \citep{bennett:2006a, bhattacharya:2023a}, but there have been no binary lens star pairs to attempt this measurement until now. We detect a weak (${\sim}1.3\sigma$) signal at the location of the lens, which may indicate a very wide orbit for the two lens stars and thus support the circumstellar scenario for the planetary companion. However our measurement is not significant enough for us to place any reasonable constraints on the 2D projected separation of the two lens stars in the 2025 epoch.\\
\indent Finally, although there remains ambiguity between the circumbinary and circumstellar interpretation for the planet, our measurement of the total lens system brightnesses (Table \ref{table:dual-phot}) and lens-source separations (Table \ref{table:epoch_seps}) enable us to predict the level of RV precision required by next-generation spectrographs on ELTs. If the two lens stars are in a close orbit and have $i \ge 15\degree$, then future RV measurements with precisions of ${\sim}1$ km/s should be able to detect this signal which would support the circumbinary planet scenario. A non-detection of the RV semi-amplitude signal would imply that the orbital inclination is nearly face-on or that the two lens stars are at a much wider orbit and further support the circumstellar planet scenario.
\vspace{2mm}

\section*{Acknowledgments}
\noindent Support for program GO-17834 was provided by NASA through a grant from STScI, which is operated by AURA, Inc., under NASA contract NAS 5-26555. The HST data presented in this article were obtained from the Mikulski Archive for Space Telescopes (MAST) at the Space Telescope Science Institute. The specific observations analyzed can be accessed via \dataset[doi: 10.17909/hzwv-qf56]{https://doi.org/10.17909/hzwv-qf56}. The Keck Telescope observations and data analysis were supported by a NASA Keck PI Data Award, 80NSSC18K0793, administered by the NASA Exoplanet Science Institute. Data presented herein were obtained at the W. M. Keck Observatory from telescope time allocated to the National Aeronautics and Space Administration through the agency's scientific partnership with the California Institute of Technology and the University of California. The Observatory was made possible by the generous financial support of the W. M. Keck Foundation. The authors wish to recognize and acknowledge the very significant cultural role and reverence that the summit of Maunakea has always had within the indigenous Hawaiian community. We are most fortunate to have the opportunity to conduct observations from this mountain. This work was supported by the University of Tasmania through the UTAS Foundation and the endowed Warren Chair in Astronomy and the ANR COLD-WORLDS (ANR-18-CE31-0002).

\noindent \textit{Software}: DAOPHOT$-$MCMC \citep{terry:2021a}, eesunhong \citep{bennett:1996a, bennett:2024a}, genulens \citep{koshimoto:2021a}, KAI \citep{lu:2022a}, Matplotlib \citep{hunter:2007a}, MORIA \citep{bhadra:2026a}, Numpy \citep{harris:2020array}

\bibliographystyle{aasjournal}
\bibliography{Terry_ob06284.bib}

\end{document}